\documentclass{kapproc} 

\usepackage{latexsym}
\usepackage{epsfig}
\usepackage{epsf}
\usepackage{graphicx}

\let\footnote\savefootnote

\kluwerbib

\def\deg{\ifmmode ^\circ                
         \else $^\circ$
         \fi
         \hskip -0.1truecm}
\def\degd#1.#2{                         
               \ifmmode {#1^{\hskip 0.05em\circ}\hskip-0.42em.\hskip0.08em#2}
               \else {#1$^{\hskip 0.05em\circ}\hskip-0.42em.\hskip0.08em$#2}
               \fi
              }
\def\mind#1.#2{                         
               \ifmmode {#1^{\hskip 0.05em\prime}\hskip-0.35em.\hskip0.05em#2}
               \else {#1$^{\hskip 0.05em\prime}\hskip-0.35em.\hskip0.05em$#2}
               \fi
              }
\def\secd#1.#2{                         
               \ifmmode {#1^{\prime\prime}\hskip-0.46em.\hskip0.12em#2}
               \else {#1$^{\prime\prime}\hskip-0.46em.\hskip0.12em$#2}
               \fi
              }
\def\timsecd#1.#2{                      
                  \ifmmode {#1^{\rm s}\hskip-0.39em.\hskip0.08em#2}
                  \else {$#1^{\rm s}\hskip-0.39em.\hskip0.08em#2$}
                  \fi
                 }
\def\hms#1h#2m#3s{                      
                  \relax
                  \ifmmode #1^{\rm h}\,#2^{\rm m}\,#3^{\rm s}
                  \else \hbox{$#1^{\rm h}\,#2^{\rm m}\,#3^{\rm s}$}
                  \fi
                 }
\def\dms#1d#2m#3s{                      
                  \relax
                  \ifmmode #1^\circ\,#2^{\prime}\,#3^{\prime\prime}
                  \else \hbox{$#1^\circ\,#2^{\prime}\,#3^{\prime\prime}$}
                  \fi
                 }
\def\dmsd#1d#2m#3.#4s{                  
                      \relax
                      \ifmmode #1^\circ\,#2^{\prime}\,#3^{\prime\prime}
                               \hskip-0.46em.\hskip0.12em#4
                      \else \hbox{$#1^\circ\,#2^{\prime}\,#3^{\prime\prime}
                            \hskip-0.46em.\hskip0.12em#4$}
                      \fi
                     }
\def\hm#1h#2m{                          
              \relax
              \ifmmode #1^{rm h}\,#2^{\rm m}
              \else \hbox{$#1^{\rm h}\,#2^{\rm m}$}
              \fi
             }
\def\dm#1d#2m{                          
              \relax
              \ifmmode #1^\circ\,#2^{\prime}
              \else \hbox{$#1^\circ\,#2^{\prime}$}
              \fi
             }
\def\hmsd#1h#2m#3.#4s{                  
                      \relax
                      \ifmmode #1^{\rm h}\,#2^{\rm m}\,#3^{\rm s}
                               \hskip-0.39em.\hskip0.08em#4
                      \else \hbox{$#1^{\rm h}\,#2^{\rm m}\,#3^{\rm s}
                            \hskip-0.39em.\hskip0.08em#4$}
                      \fi
                     }
\def\hmd#1h#2.#3m{                  
                  \relax
                  \ifmmode #1^{\rm h}\,#2^{\rm m}
                           \hskip-0.55em.\hskip0.22em#3
                  \else \hbox{$#1^{\rm h}\,#2^{\rm m}
                        \hskip-0.55em.\hskip0.22em#3$}
                  \fi
                 }
\def\mg{\relax                          
        \ifmmode ^{\rm m}
        \else $^{\rm m}$
        \fi
       }
\def\mgd#1.#2{                          
              \relax
              \ifmmode #1^{\rm m}
                       \hskip-0.55em.\hskip0.22em#2
              \else \hbox{#1$^{\rm m}
                    \hskip-0.55em.\hskip0.22em$#2}
              \fi
             }

\def\la{\mathrel{\hbox{\rlap{\hbox{\lower4pt\hbox{$\sim$}}}\hbox{$<$}}}}
\def\ga{\mathrel{\hbox{\rlap{\hbox{\lower4pt\hbox{$\sim$}}}\hbox{$>$}}}}

\def\unitspace{\;}                      

\def\un#1{\ifmmode \unitspace\mbox{\rm #1} 
          \else $\unitspace$#1
          \fi}
\def\pun#1#2{\ifmmode \unitspace\mbox{\rm #1}^{#2} 
             \else $\unitspace$#1$^{#2}$
             \fi}

\def\kms{\un{km}\pun{s}{-1}}          
\def\Lsun{\ifmmode \un{L}_{\odot}     
          \else $\un{L}_{\odot}$
          \fi}
\def\Msun{\ifmmode \un{M}_{\odot}     
          \else $\un{M}_{\odot}$
          \fi}
\def\mum{\ifmmode \unitspace\mu\mbox{\rm m} 
         \else $\unitspace\mu$m
         \fi}
\def\pyr{\pun{yr}{-1}}                
\def\sqarcsec{\ifmmode \unitspace\Box''    
              \else $\unitspace\Box''$     
              \fi} 

\def\Bp{\relax                            
        \ifmmode B_{||}                   
        \else $B_{||}$
        \fi}
\def\Bt{\relax                            
        \ifmmode B\!_{\perp}              
        \else $B\!_{\perp}$               
        \fi}
\def\Gcr{\relax                           
         \ifmmode \Gamma\!_{\rm cr}       
         \else $\Gamma\!_{\rm cr}$
         \fi}
\def\ICII{\relax                          
          \ifmmode I_{[\CII]}             
          \else $I_{[\CII]}$
          \fi}
\def\LHtwo{\relax                                 
           \ifmmode L_{\mbox{\rm\scriptsize H}_2} 
           \else $L_{\mbox{\rm\scriptsize H}_2}$  
           \fi}
\def\LLya{\relax                          
          \ifmmode L_{{\rm Ly}\,\alpha}   
          \else $L_{{\rm Ly}\,\alpha}$
          \fi}
\def\MHtwo{\relax                                 
           \ifmmode M_{\mbox{\rm\scriptsize H}_2} 
           \else $M_{\mbox{\rm\scriptsize H}_2}$  
           \fi}
\def\MHtwodot{\relax                                       
              \ifmmode \dot{M}_{\mbox{\rm\scriptsize H}_2} 
              \else $\dot{M}_{\mbox{\rm\scriptsize H}_2}$  
              \fi}                                         
\def\Mstardot{\relax                      
              \ifmmode \dot{M}_{\ast}     
              \else $\dot{M}_{\ast}$      
              \fi}
\def\nHI{\relax                                      
         \ifmmode n_{\mbox{\scriptsize\rm H\,\sc I}} 
         \else $n_{\mbox{\scriptsize\rm H\,\sc I}}$
         \fi}
\def\nHtwo{\relax                                
           \ifmmode n_{{\mbox{\scriptsize H}}_2} 
           \else $n_{{\mbox{\scriptsize H}}_2}$  
           \fi}
\def\rhostardot{\relax                         
                \ifmmode \dot{\rho}_{\ast}     
                \else $\dot{\rho}_{\ast}$      
                \fi}
\def\rhoZdot{\relax                          
             \ifmmode \dot{\rho}_{\rm Z}     
             \else $\dot{\rho}_{\rm Z}$      
             \fi}

\def\sou#1#2{\relax                       
             \ifmmode {\rm #1}\,{\rm #2}  
             \else #1$\,$#2
             \fi}

\def\qu#1#2{\relax                          
            \ifmmode #1_{\rm #2}            
            \else $#1_{\rm #2}$
            \fi}

\def\CO#1{\ifnum#1=0                    
           \ifmmode \mbox{\rm CO}
           \else {\rm CO}
           \fi
          \else
           \ifnum#1<15
            \ifmmode ^{#1}\mbox{\rm CO}
            \else $^{#1}${\rm CO}
            \fi
           \else
            \ifmmode \mbox{\rm C}^{#1}\mbox{\rm O}
            \else {\rm C}$^{#1}${\rm O}
            \fi
           \fi
          \fi}

\def\COp{\ifmmode \mbox{\rm CO}^+           
         \else {\rm CO}$^+$                 
         \fi}

\def\CS#1{\ifnum#1=0                    
           \ifmmode \mbox{\rm CS}
           \else {\rm CS}
           \fi
          \else
           \ifnum#1<15
            \ifmmode ^{#1}\mbox{\rm CS}
            \else $^{#1}${\rm CS}
            \fi
           \else
            \ifmmode \mbox{\rm C}^{#1}\mbox{\rm S}
            \else {\rm C}$^{#1}${\rm S}
            \fi
           \fi
          \fi}

\def\HCOp{\ifmmode \mbox{\rm HCO}^+          
          \else {\rm HCO}$^+$                
          \fi}
\def\Hthreep{\ifmmode \mbox{\rm H}_3^+         
             \else {\rm H}$_3^+$               
             \fi}

\def\Htwo{\ifmmode \mbox{\rm H}_2              
          \else {\rm H}$_2$                    
          \fi}

\def\HtwoO{\ifmmode \mbox{\rm H}_2\mbox{\rm O} 
           \else {\rm H}$_2${\rm O}            
           \fi}

\def\ion#1#2{\ifmmode \mbox{{\rm #1}}\,\mbox{{\sc #2}} 
        \else {\rm #1}$\,${\sc #2}
        \fi}

\def\rec#1#2{\if#2a                            
              \ifmmode \mbox{{\rm #1}}\alpha   
              \else {\rm #1}$\alpha$
              \fi
             \fi
             \if#2b
              \ifmmode \mbox{{\rm #1}}\beta
              \else {\rm #1}$\beta$
              \fi
             \fi
             \if#2g
              \ifmmode \mbox{{\rm #1}}\gamma
              \else {\rm #1}$\gamma$
              \fi
             \fi}

\def\Lya{\rec{Ly}{a}}                          
\def\Ha{\rec{H}{a}}                            

\newcommand{\figref}[1]{Fig.~\protect\ref{#1}}

\newcommand{\eqref}[1]{Eq.~$\left(\protect\ref{#1}\right)$}

\newcommand{\secref}[1]{Sect.~\protect\ref{#1}}

\begin{document}

\articletitle{Probing the evolution of galaxies using redshifted
H$\alpha$ emission}


\author{Paul P.\ van der Werf}
\affil{Leiden Observatory, Leiden, The Netherlands}
\email{pvdwerf@strw.leidenuniv.nl}

\author{Alan F. M. Moorwood}
\affil{European Southern Observatory, Garching, Germany}
\email{amoor@eso.org}

\author{Lin Yan}
\affil{SIRTF Science Center, Caltech, Pasadena, CA, U.S.A.}
\email{lyan@ipac.caltech.edu}

\makeatletter
\renewcommand{\@makefnmark}{\mbox{\ }}
\makeatother
 
\renewcommand{\thefootnote}{}

\footnote{invited review, to appear in {\it Processings of the 2nd
Hellenic Cosmology Workshop}, eds.\ M.\ Plionis et al. (Kluwer)}

\vspace{-0.5cm}

\begin{abstract}
In this paper we review the present status and implications of $\Ha$
surveys at various redshifts. With the advent of sensitive wide-format
near-infrared detectors on large telescopes, deep and extensive $\Ha$
surveys are now feasible to redshift $z\sim2.5$. The cosmic star formation
history can therefore be traced out to this redshift using $\Ha$
alone, avoiding complications arising from the comparison of different
tracers at different redshifts. The $\Ha$ surveys to date confirm the
rapid increase in luminosity density from $z=0$ out to $z=1$, and
show that this increase flattens off at higher redshifts, remaining
approximately constant out to at least $z\sim2.2$.
We also discuss the prospects for determining the masses of high
redshift galaxies based on emission lines. A set of high-quality $\Ha$
rotation curves of samples of disk galaxies at a number of different
redshifts would allow a study of the evolution of the Tully-Fisher
relation and address fundamental issues in disk galaxy
formation. Such a program remains challenging even with present-day
large telescopes.
\end{abstract}


\section{Introduction}
\label{sec.introduction}

When did the stellar populations of present-day galaxies form? This
question is now beginning to be addressed by observations with modern
telescopes in virtually all regions of the electromagnetic spectrum.
Moderate redshift ($z<1$) surveys consistently indicate a cosmic star
formation rate density (SFRD, the stellar mass formed per unit of time
and per unit of comoving volume) 
$\rhostardot$ that strongly increases
with redshift out to $z\sim1$ (\cite{Hogg01}). The SFRD at $z>1$ is
however a more hotly debated issue. Estimates based on surveys of Lyman
break galaxies (e.g., \cite{Madauetal96}) sample rest-frame UV
radiation and require a large and uncertain extinction correction
(e.g., \cite{Pettinietal98}; \cite{Steideletal99}).
Submillimetre surveys, tracing reradiation by dust of absorbed
starlight (e.g., \cite{Hughesetal98}; \cite{Blainetal99b})
consistently indicate an even higher SFRD at these redshifts than even
the extinction-corrected UV-determined values.
As a result, the use of different star
formation tracers at different redshifts complicates the
reconstruction of the cosmic star formation history and it is therefore
highly advantageous to use the same star
formation tracer at all redshifts. The $\Ha$
line is the natural choice for this tracer, since with the
development of large area, high quality near-infrared arrays, this
line can now be observed in the $J$, $H$ and $K$-bands, where
redshifts of approximately 1, 1.5 and 2.2 respectively, are
accessible.

Traditionally, emission-line surveys for high-$z$ star forming
galaxies have targeted the $\rec{Ly}{a}$ line, which moves into the
optical regime for $z=1.9-7.0$. However, since $\rec{Ly}{a}$ is
resonantly scattered, even very small quantities of dust will
effectively suppress the line (Charlot \& Fall 1991, 1993;
\cite{ChenNeufeld94}).  Spectroscopic observations of star forming
$z>3$ Lyman break galaxies
(\cite{Steideletal96a}) confirm that $\rec{Ly}{a}$ is not a good
tracer of star formation in high-$z$ galaxies. These galaxies form
stars at rates of $\sim10\Msun\pun{yr}{-1}$ as derived from their
restframe UV properties (uncorrected for extinction), but
$\rec{Ly}{a}$ is absent (or in {\it absorption}) in more than 50\% of
the cases, while in most of the remaining objects the line is faint.
Thus, while $\rec{Ly}{a}$ searches can be used to find 
high-$z$ star forming galaxies (e.g., \cite{HuMcMahon96}; \cite{Huetal98};
\cite{CowieHu98}; \cite{Manningetal00}; \cite{Sternetal00};
\cite{Kudritzkietal00}; \cite{Rhoadsetal00}; \cite{Steideletal00}),
such surveys cannot provide reliable star formation rates (SFRs)
for the galaxies that are detected.
\nocite{CharlotFall91} \nocite{CharlotFall93}

This complication is avoided by searching for $\rec{H}{a}$ emission in
stead of $\rec{Ly}{a}$. The $\rec{H}{a}$ line is not resonantly
scattered and thus much less sensitive to the effects of small amounts
of dust. In addition, the broad-band extinction at $\rec{H}{a}$
($6563\un{\AA}$) is much less than that at $\rec{Ly}{a}$
($1215\un{\AA}$) ($A_{{\rm Ly}\alpha}/A_{{\rm H}\alpha}=4.28$ for the
extinction curve used by \cite{Cardellietal89}).  Although the
extinction at the wavelength of $\Ha$ is still large in starburst
galaxies, the typical $\Ha$ extinction in local spiral galaxies is
only $\mgd 1.1$ (\cite{Kennicutt83} and references therein).  A recent
comparison of star formation tracers (including stellar continuum
emission as well as nebular lines) confirms that $\Ha$ is the most
robust star formation tracer, and can give results accurate to within a
factor of 2 to 3 when combined with other tracers, even if the
$\Ha$/[$\ion{N}{ii}$] blend is not resolved (\cite{CharlotLonghetti01}).
In contrast, the also widely used [$\ion{O}{ii}$] line at
$3727\un{\AA}$ is a more uncertain star formation tracer by a full
order of magnitude, because of uncertainties in excitation,
metallicity and reddening. The SFR derived from $\Ha$ also correlates
well with that estimated from the UV
(\cite{BellKennicutt01}). The systematically higher SFR derived from $\Ha$ (by
a factor of 1.5 in relatively luminous galaxies) simply
reflects the difference in
extinction between the optical and UV regimes.

In this paper we first 
review the current status of $\Ha$ surveys from $z=0$ out to
the highest redshifts where such surveys have been performed to date. Since
we only consider $\Ha$, this analysis should be free of uncertainties 
introduced by using different tracers at different redshifts. Uncertainties
in the conversion of $\Ha$ luminosity into SFR rate can be avoided by quoting
the $\Ha$ luminosity density, which can be derived directly by integrating
the $\Ha$ luminosity function.\\

In the second part of this review we address the use of $\Ha$ for
studying the kinematics of high-$z$ galaxies. While the SFRD is
expressed as a time derivative of comoving {\it mass\/} density, the
various ways of determining SFRD do not involve the measurement of
mass but of light. However, the stars producing most of the light
contain very little mass, while the stars containing most of the mass
produce little light, for a typical initial mass function (IMF). As a
result, calculating the {\it total\/} SFR from a measurement of light
(regardless of whether this is rest-frame ultraviolet light, nebular
line emission, or emission from dust) always involves the uncertain
step of assuming an IMF\null. While systematic errors in the analysis
can be mitigated by ensuring that always the same IMF is {\it
assumed}, the universality of the IMF is still an assumption and a
very debatable issue. The most direct way of avoiding this
complication is by making a dynamical mass
measurement. Spectral line measurements offer this possibility,
provided the kinematic pattern is dominated by gravitational
motion.  For this reason, $\Lya$ kinematics is normally not useful
since $\Lya$ line profiles are often affected by nongravitational
motions such as outflows.  Emission line kinematics can be assumed to
be dominated by gravitational motion if a characteristic pattern such
as a rotation curve is observed. Even then, one must be wary of
excitation effects, since lines such as [$\ion{O}{iii}$]
$5007\un{\AA}$ possibly sample only the inner parts of a galaxy, and may
not reveal the full amplitude of the rotation curve. Again, using
$\Ha$ is the best way for avoiding such biases.

Dynamical measurements of this type have the potential for probing the
evolution of the Tully-Fisher (TF) luminosity-line width relation out
to $z\sim2.5$. This issue is directly related to the physics of disk
galaxy formation. One model for the origin of the TF relation argues
that the dark halo is the sole factor determining the disk properties
(and therefore the TF relation), since the outer region of
disks, where the rotation curves are flat, are dynamically dominated by
the dark halo.  Models based on this observation predict rapid
evolution of the TF relation \hbox{(\cite{Moetal98})}. On the other hand, it
has been proposed that the TF relation is entirely the result of star
formation self-regulation in the disk interstellar medium and has
nothing to do with the halo at all (\cite{Silk97}). A set of high-quality $\Ha$
rotation curves of disk galaxies at a number of different redshifts is
needed to settle this fundamental issue from an observational point of
view.  First results and prospects in this challenging field are
reviewed in \secref{sec.rotation}.

Unless noted otherwise, in this review, all quantities that are
dependent on cosmology are quoted for a universe with
$(h,\qu{\Omega}{M},\Omega_\Lambda) = (0.5,0.5,0.0)$, where the Hubble
constant is $\qu{H}{0}=100\,h\kms\pun{Mpc}{-1}$ and $\qu{\Omega}{M}$
and $\qu{\Omega}{\Lambda}$ are the present-day density parameters of
matter and of the cosmological constant, respectively. Where
necessary, results originally quoted for other cosmologies have been
tacitly converted.

\section{The cosmic star formation history derived from H$\alpha$ surveys}
\label{sec.surveys}

\subsection{Surveys and strategies}

Observationally, the task of measuring the SFRD
at a particular redshift from $\Ha$ consists of
determining the $\Ha$ luminosity function, which is parametrized 
(\cite{Schechter76}) as
\begin{equation}
\phi(L)\,dL = \phi^*\,(L/L^*)^\alpha\,e^{-L/L^*}\,d(L/L^*)
\end{equation}
and gives the number density (per Mpc$^3$) 
of galaxies with luminosities between 
$L$ and $L+dL$. For cosmological applications, 
{\it comoving\/} number densities
(i.e., number of galaxies per comoving Mpc$^3$) are computed.
In practice, the number density of galaxies is binned per 0.4 interval in 
$\log L$, which can be related to the standard form by
\begin{equation}
\Phi(\log L)\,{d\,\log L\over 0.4} = \phi(L)\,dL.
\end{equation}
The total luminosity density is found by integrating the luminosity function 
over all luminosities:
\begin{equation}
\rho_L = \int L\,\phi(L)\, dL = \phi^*\,L^*\,\Gamma(2+\alpha),
\end{equation}
where $\Gamma$ denotes the gamma function. For a determination of the total
luminosity density it is therefore sufficient to determine the three parameters
$\phi^*$, $L^*$ and $\alpha$ of the luminosity function. Note that the
integrated luminosity density is dominated by low-luminosity objects
if $\alpha$ approaches $-2$.

Surveys for emission line objects can be of three types (cf.,
\cite{Djorgovski92}). Slitless spectroscopic surveys sample large
volumes of space but can only achieve interesting depths when carried
out from space. Ground-based surveys can be either serendipitous
long-slit surveys (covering a large redshift range but a very limited
solid angle) or narrow-band surveys (sampling a large area but in a
very narrow redshift interval).  Since the background in the near-IR
windows shortward of about $2.2\mum$ is dominated by OH emission
lines, the narrow-band survey technique is the ideal approach in this
spectral region for ground-based observations.  This technique
involves deep imaging in a suitable narrow-band filter, complemented
with broad-band imaging; sources with excess flux in the narrow-band
filter are emission line candidates in a redshift interval determined
by the narrow-band filter passband (this strategy has been analyzed in
detail by \cite{MannucciBeckwith95}). Narrow-band surveys of
$\rec{H}{a}$ emission at $z>2$ were initially unsuccesful
(\cite{Thompsonetal94}; \cite{PahreDjorgovski95}; \cite{Bunkeretal95};
\cite{Collinsetal96}; \cite{Thompsonetal96}) due to lack of depth
and/or coverage (with the exception of one object at $z=2.43$ reported by
\cite{Beckwithetal98}).  In contrast, more extensive $\Ha$ surveys
targeted at volumes containing known damped $\rec{Ly}{a}$ and metal-line
absorbers, quasars and radio galaxies resulted in the identification
of a significant number of candidate objects
(\cite{Mannuccietal98}; \cite{Teplitzetal98}; 
\cite{VanDerWerfetal00}). However, since none of these surveys has been
succesfully followed up with spectroscopic confirmation, these results
must so far be considered tentative. More
fundamentally, since these were {\it targeted\/} surveys, centred at
particular ``marker'' objects of known redshift, they are most likely
biased to overdense regions and cannot be used to derive a star
formation rate density that is valid for the universe on global
scales.

\begin{figure}[t]
\vskip.2in
\mbox{\epsfxsize=12cm \epsffile{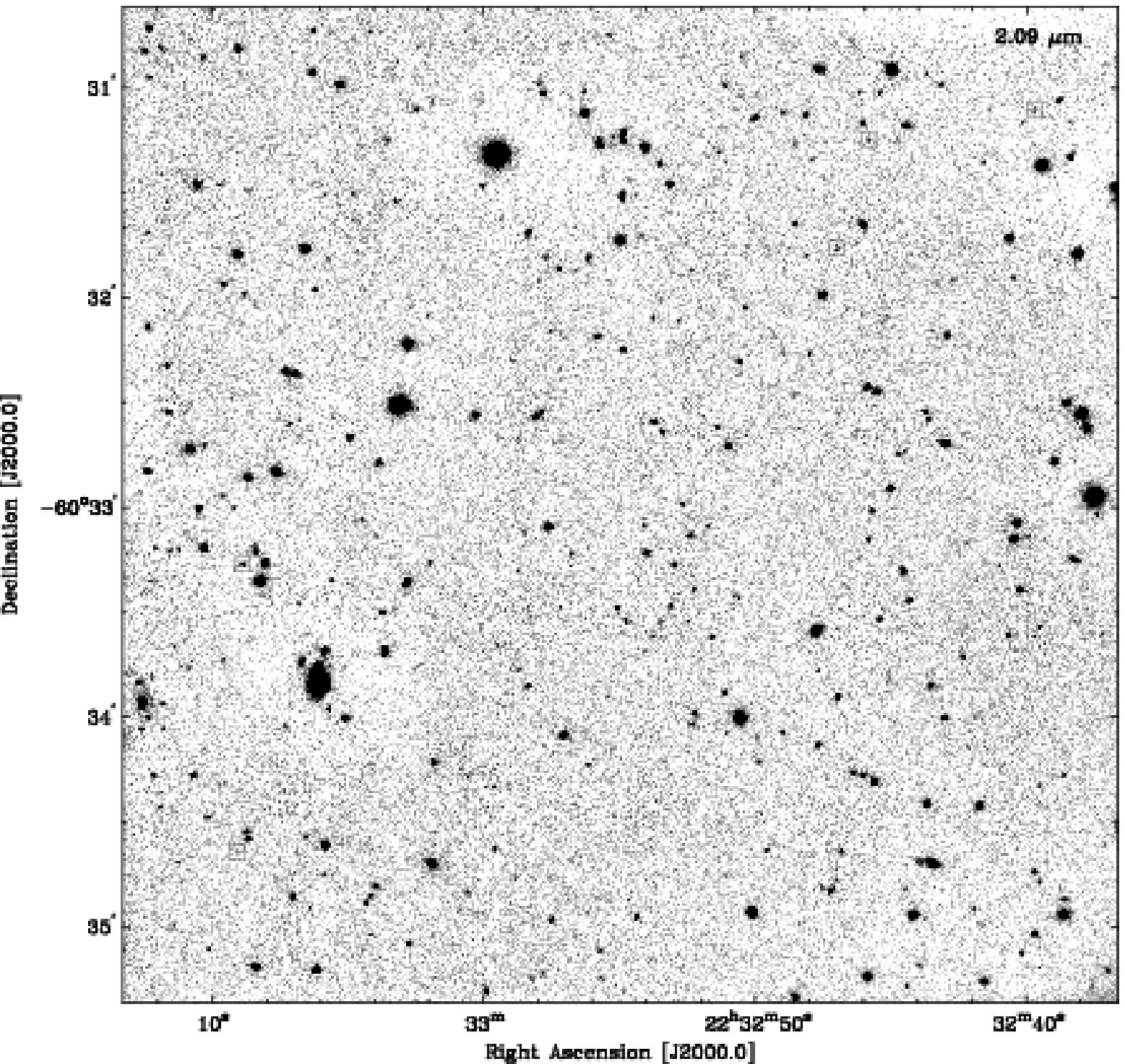}}
\letteredcaption{a}{Narrow-band image at $2.09\mum$
centred on the HDFS WFPC2 field. 
The field size is $5'\times5'$ with North at the top and East to the left. 
The squares identify spectroscopically confirmed emission-line sources 
(from \protect{\cite{Moorwoodetal00a}}).}
\label{fig.WFPC2images}
\end{figure}

\begin{figure}[t]
\vskip.2in
\mbox{\epsfxsize=12cm \epsffile{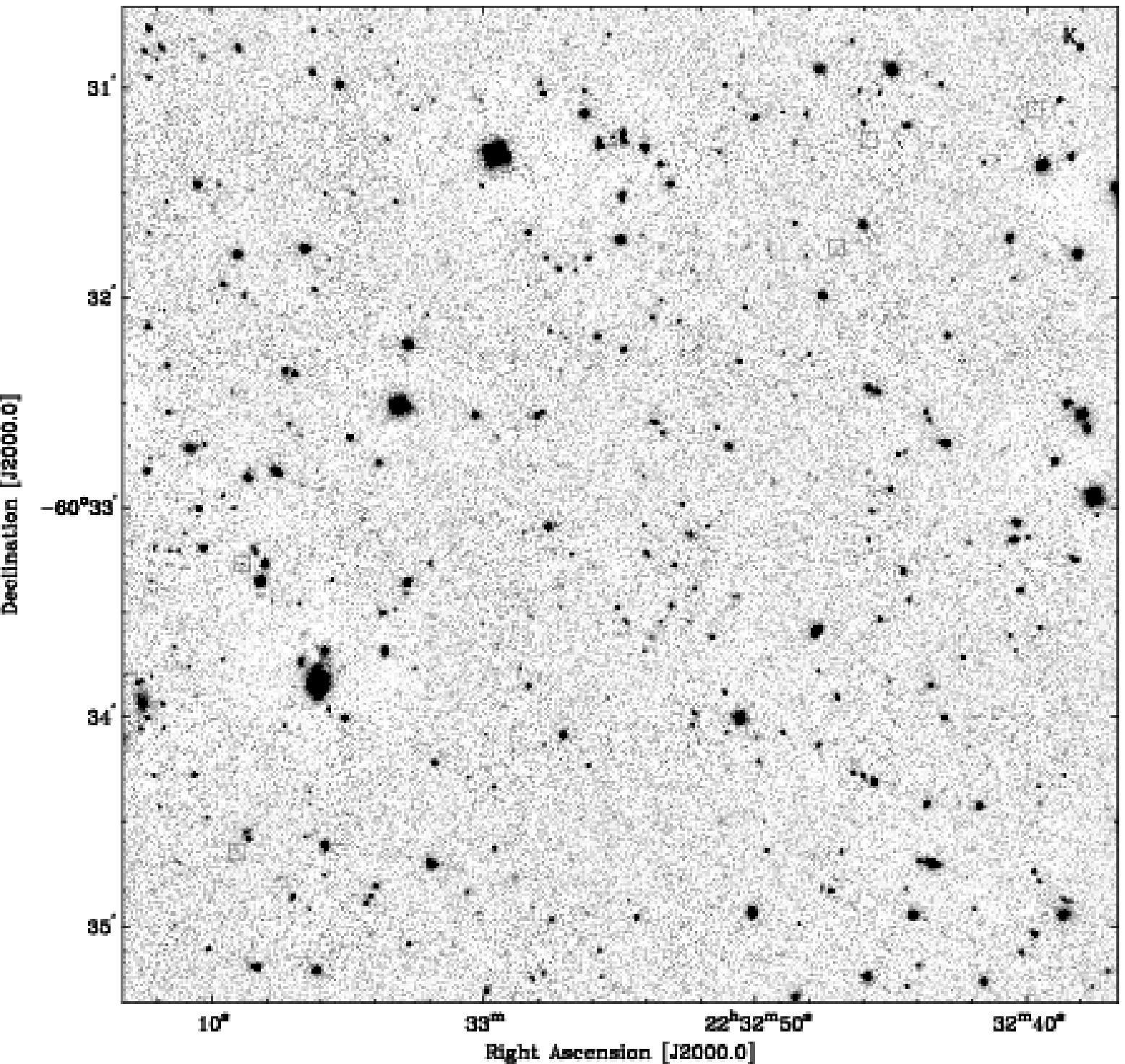}}
\letteredcaption{b}{Broad-band $\qu{K}{s}$ 
image corresponding to the \protect{$2.09\mum$} narrow-band
image of the HDFS WFPC2 field (from \protect{\cite{Moorwoodetal00a}}).}
\end{figure}

\subsection{Blank field H$\alpha$ surveys}

\subsubsection{Surveys at $z\la1$}
\label{sec.z0}

The $\Ha$ luminosity density of the local ($z<0.045$) universe has
been determined from the Universidad Complutense de Madrid objective
prism survey (\cite{Gallegoetal95}). At somewhat higher redshift, the
$\Ha$ luminosity density derived from a $\left< z \right>\sim0.2$
sample taken from the $I$-band selected Canada France Redshift Survey
(CFRS) is about a factor of 2 higher (\cite{TresseMaddox98}). A small
survey of 13 CFRS galaxies at $z\approx 0.9$ produced an $\Ha$
luminosity density approximately 10 times higher than that at $z\sim0$
\mbox{(\cite{Glazebrooketal99})}.  All of these results include a small
(approximately 1 magnitude) correction for extinction, and a
correction for the effects of the [$\ion{N}{ii}$] lines on the derived
$\Ha$ flux. The magnitude of this increase is in excellent agreement
with the increase in cosmic luminosity density measured in the
rest-frame ultraviolet (\cite{Lillyetal96}) and in radio continuum
emission (\cite{Haarsmaetal00}).
The redshift region out to $z\sim1$ is thus characterized by a strong
increase in luminosity density with lookback-time. While the precise
values of the corresponding star formation rate densities are
uncertain due to the complications discussed in
\secref{sec.introduction}, the increase out to $z\sim1$, when measured
in one particular tracer, is now well documented and is quantitatively
consistent between various different tracers (\cite{Hogg01}).

It should be emphasized that existing $\Ha$ surveys at $z\la1$ 
are limited in both width
(leading to poor statistics, especially at the high luminosity end)
and depth (leading to a poor determination of the faint-end slope of
the luminosity function). Indeed, a recent deeper narrow-band survey
for $\Ha$ at redshifts of approximately 0.08, 0.2 and 0.4
(\cite{JonesBlandHawthorn01}) suggests the need for an upward revision
of the local $\Ha$ luminosity density, mainly as a result of a higher
detection rate of faint galaxies. This result is in agreement with an
analysis of the KPNO International Spectroscopic Survey
(\cite{Gronwall99}), based on 1126 $\Ha$ emitters at $z<0.085$
detected using the objective prism technique.

\subsubsection{Surveys at $1\la z\la2$}

The redshift range between redshifts 1 and 2 is of fundamental
interest since it is in this range that the cosmic star formation
history starts to turn over from a fairly high level of star formation
rate density (perhaps to first order constant with redshift at $z\ga1$)
into the rapid decrease discussed in
the previous section. The shape, origin, and redshift of this turnover
are not well known. This lack of knowledge is partly explained by the
lack of suitable redshift tracers in the optical region for redshifts
between 1 and 2. In addition, at $z\ga1$ the $\Ha$ line has moved into
the near-infrared regime and sensitive large-format near-infrared
cameras and spectrographs at large telescopes have only recently become
available.

The most important $\Ha$ surveys at $1\la z\la 2$ to date are the two
slitless grism surveys carried out with NICMOS on the Hubble Space
Telescope.  The first of these surveys (\cite{McCarthyetal99}) covered
approximately $64\Box'$ and yielded an $\Ha$ luminosity function
(\cite{Yanetal99}) with $L^*(\Ha)$ exceeding the $z=0$ value
(\cite{Gallegoetal95}) by a factor of 7, implying significant
luminosity evolution. The second survey (\cite{Hopkinsetal00}), covered
only $4.4\Box'$ but extended the $\Ha$ luminosity function towards
fluxes fainter by a factor of 2, pinning down the faint-end slope of
the luminosity function at $\alpha=-1.6$. In total these surveys
produced 70 $\Ha$ emitters between $z=0.7$ and 1.9. A significant
fraction of these have been confirmed as $\Ha$ emitters by
spectroscopy of other emission lines.  Since the two grism surveys
define the $\Ha$ luminosity function at $\left< z \right>\sim1.3$ over
a significant range in luminosity, the implied $\Ha$ luminosity
density is well-determined. The resulting value 
is $3.6\cdot10^6\,\Lsun\pun{Mpc}{-3}$, without any correction for
extinction. It is remarkable that this value is 15\% {\it smaller\/}
than the $z=0.9$ value (\cite{Glazebrooketal99}), 
and the steep rise in luminosity
density from $z=0$ to 1 therefore appears to flatten off at
$z>1$.

\subsubsection{Surveys at $z>2$}

\begin{figure}[t]
\vskip.2in
\mbox{\epsfxsize=14cm \epsffile{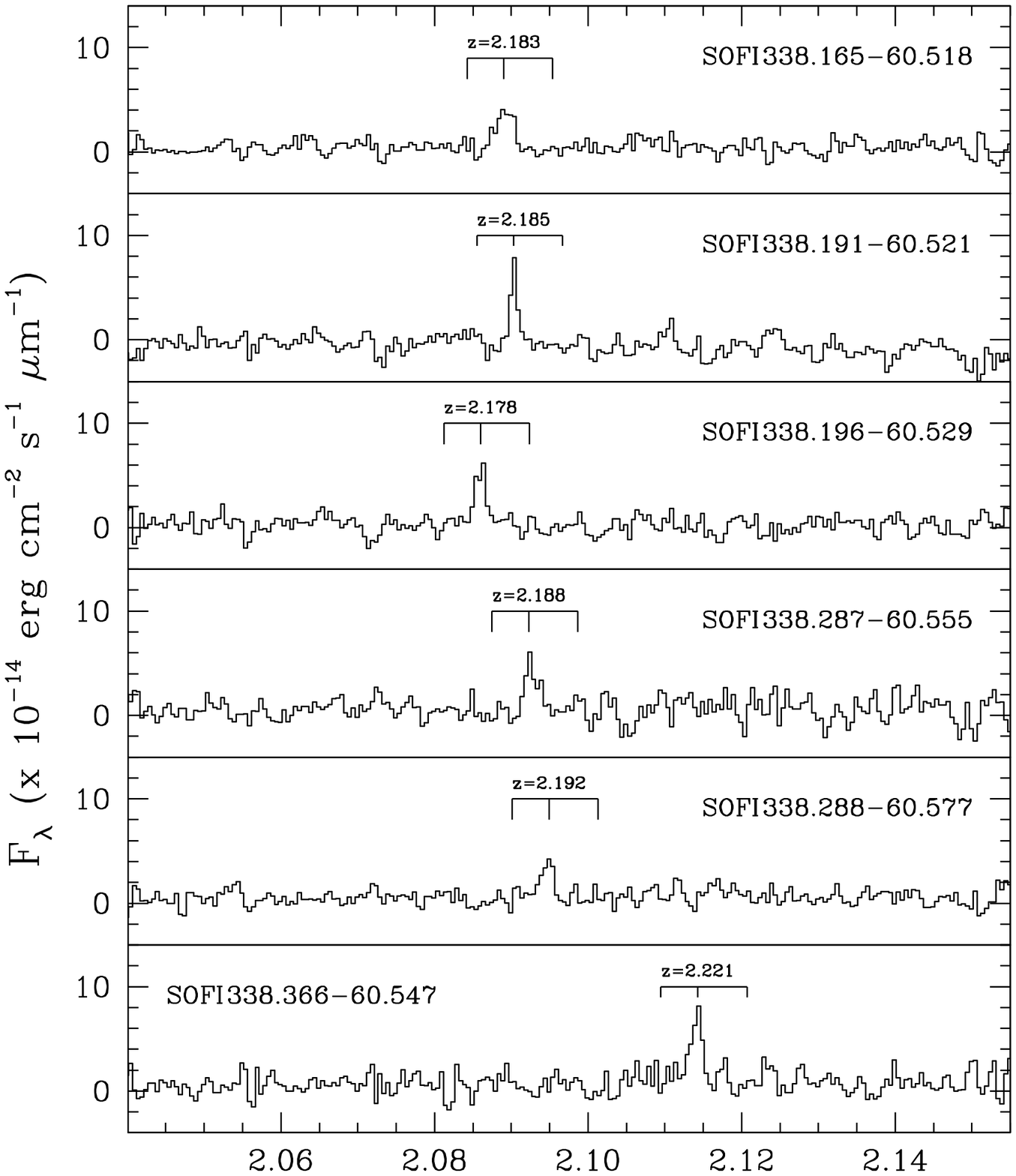}}
\caption{ISAAC spectra of emission line galaxies from the HDFS narrow-band
imaging survey (\protect{\cite{Moorwoodetal00a}}). 
The markers under the redshift
labels show the expected wavelengths of the [$\ion{N}{ii}$] 
(6548 and $6584\un{\AA}$) lines assuming the detected line is $\Ha$.}
\label{fig.spectra}
\end{figure}

The highest redshift survey for $\Ha$ emitters used the SOFI
near-infrared camera on the ESO New Technology Telescope to target a
$100\Box'$ area in the region of the Hubble Deep Field South (HDFS), in
a narrow ($\Delta z\approx 0.04$) redshift interval at $z\approx 2.2$,
using the narrow-band imaging technique (\cite{Moorwoodetal00a}). This
redshift interval was chosen because it redshifts $\Ha$ to $2.09\mum$,
which is a spectral region relatively free of bright OH lines and not
strongly affected by the thermal background that becomes dominant at
somewhat longer wavelengths. The importance of this survey lies in the
fact that it is the first succesful blank-field survey of this type
and at this redshift, and the only one with substantial spectroscopic
confirmation. One field from this survey is shown in Fig.~1. 
It reveals 5 emission line objects, all of which
are spectroscopically confirmed. A set of confirmation spectra,
obtained with ISAAC at the ESO Very Large Telescope, is shown in
\figref{fig.spectra}. The number of detections from this survey
matches precisely the number expected under the assumption that the
$\Ha$ luminosity function at $z\sim2.2$ is identical to that at
$z\sim1.3$. 
The number density of of $\Ha$ emitters at $z\sim 2.2$ is
also comparable to that of Lyman break galaxies with similar SFRs
at $z=3.0-3.5$ (\cite{Steideletal96a}).
These results provide further evidence for a constant SFRD
at $z>1$. 

\subsection{The evolution of the H$\alpha$ luminosity density and star formation in galaxies}

Comparison of the luminosity functions at $z\sim0$
(\cite{Gallegoetal95}) and $z\sim1.3$ (\cite{Yanetal99}) and higher
shows that both luminosity evolution and density evolution are
implied. However, luminosity evolution dominates (a factor 7 increase
in $L^*$ from $z=0$ to $z=1.3$ compared to a factor 2.6 increase in
$\phi^*$). Indeed, in the $z\sim2.2$ sample the implied SFRs are 20 to
$35\Msun\pyr$ (for the $\Ha$/SFR conversion factor of
\cite{Kennicutt98}, which is appropriate for continuous star formation 
and a Salpeter IMF, at solar metallicity). These values are
significantly higher than in typical nearby galaxies, but lower than
in extreme starburst galaxies, such as are found in submillimetre
surveys (e.g., \cite{Ivisonetal00}). This result suggests that, while the
submillimetre measurements select extreme starburst objects which may
plausibly be identified with the formation of present-day spheroids, 
the $\Ha$ measurements select a less extreme mode of star
formation. It may then be speculated that the $\Ha$ measurements trace
the buildup of disks. Dynamical measurement 
(cf., \secref{sec.rotation}) may shed light on this hypothesis.

\begin{figure}[t]
\vskip.2in
\mbox{\epsfxsize=6cm\epsffile{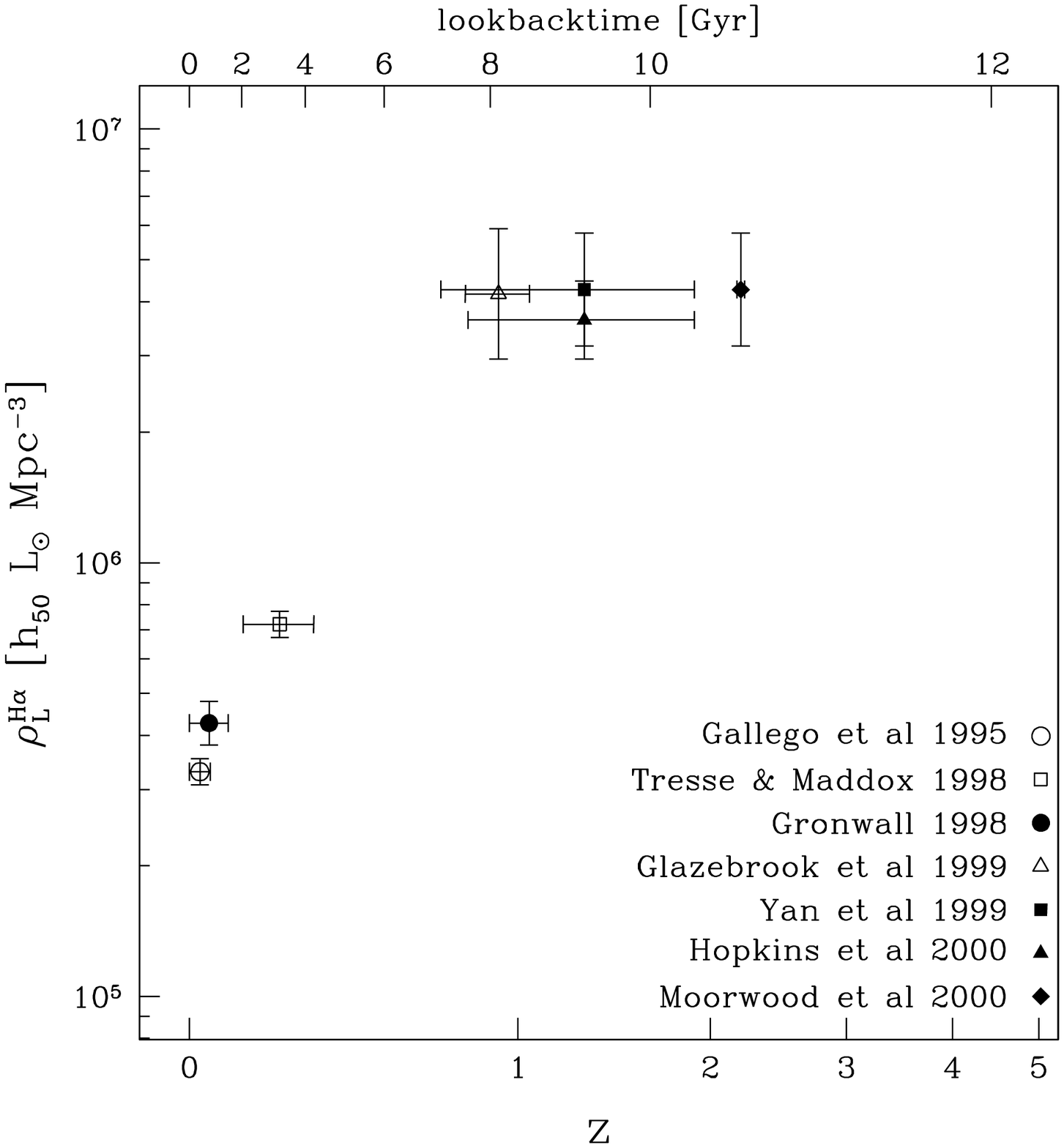} 
\epsfxsize=6cm\epsffile{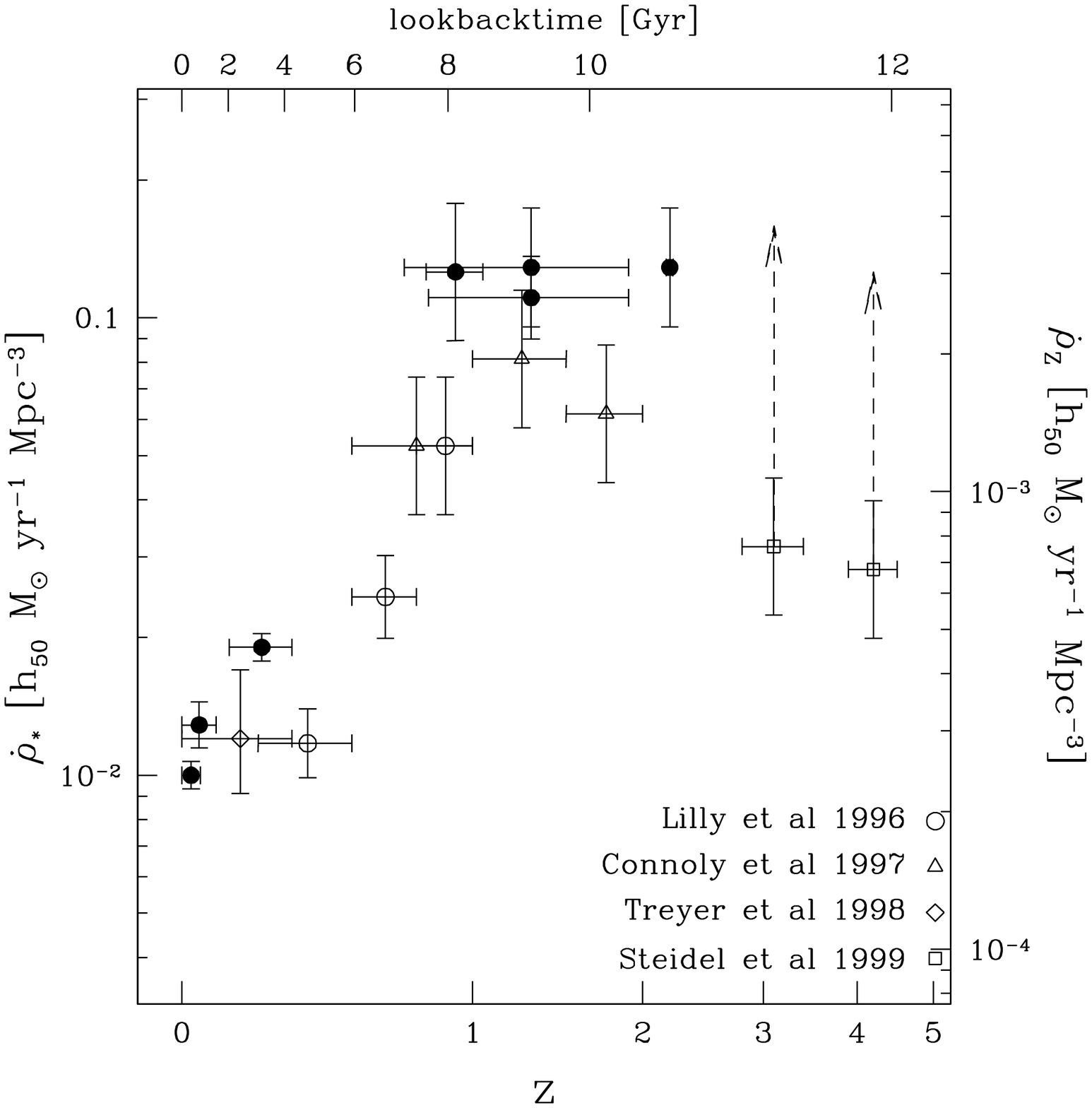}}
\caption{The evolution of the cosmic $\Ha$ luminosity density (lefthand
panel) and SFRD (righthand panel, calculated using the $\Ha$/SFR
conversion factor of \cite{Kennicutt98}). In the
lefthand panel, open symbols denote measurements that include an
extinction correction based on the Balmer decrement. In the right-hand
panel, filled symbols denote the $\Ha$-based measurements while open
symbols are based on rest-frame UV radiation; the arrows on the 2
high-redshift points in this panel indicate the magnitude of the
estimated extinction correction. In the righthand panel, the righthand
vertical axis denotes metal production rate density, which is less
sensitve to the assumed IMF than star formation rate density (\cite{Madauetal96}).
}
\label{fig.SFH}
\end{figure}

The evolution of the $\Ha$ luminosity density and the implied SFRD
history are summarized in
\figref{fig.SFH}. The $\Ha$ results give a consistent
picture of evolution out to $z\sim2.2$. Without extinction
corrections, the $\Ha$ surveys imply higher SFRDs than the UV-based
surveys, but the discrepancy is typically only a factor 
of 2 (\figref{fig.SFH}). It is interesting that the extinction
corrections proposed for Lyman break galaxies at higher redshifts lead
to an SFRD very similar to that implied by the $\Ha$ surveys (see
\figref{fig.SFH}); it should be noted however, that these corrections
are not very certain.

\section{Dynamical mass measurements at high redshifts using H$\alpha$}
\label{sec.rotation}

The low-redshift TF relation derived from $\Ha$ rotation
curves is well-defined (\cite{Courteau97}). Attempts to detect evolution
in the zero-point of the TF relation 
using $\Ha$ measurements of higher redshift
samples are so far inconclusive: while 
$\Ha$ rotation curves out to $z\sim1$
implied only about $\mgd 0.4$ evolution in $B$ (Vogt et al., 1996, 1997),
a study of [$\ion{O}{ii}$] rotation
curves at $\left< z \right> = 0.35$ suggested luminosity evolution by
$1.5-2$ magnitudes in $B$ (\cite{SimardPritchet98}). These results may
be reconciled if in the [$\ion{O}{ii}$] data the rotation curves are
not sampled all the way out to the flat part. The discrepancy
underlines the need for deep uniform $\Ha$ rotation curves for
projects of this type.\\
\nocite{Vogtetal96}
\nocite{Vogtetal97}

\begin{figure}[t]
\vskip.2in
\includegraphics[angle=270,width=12cm]{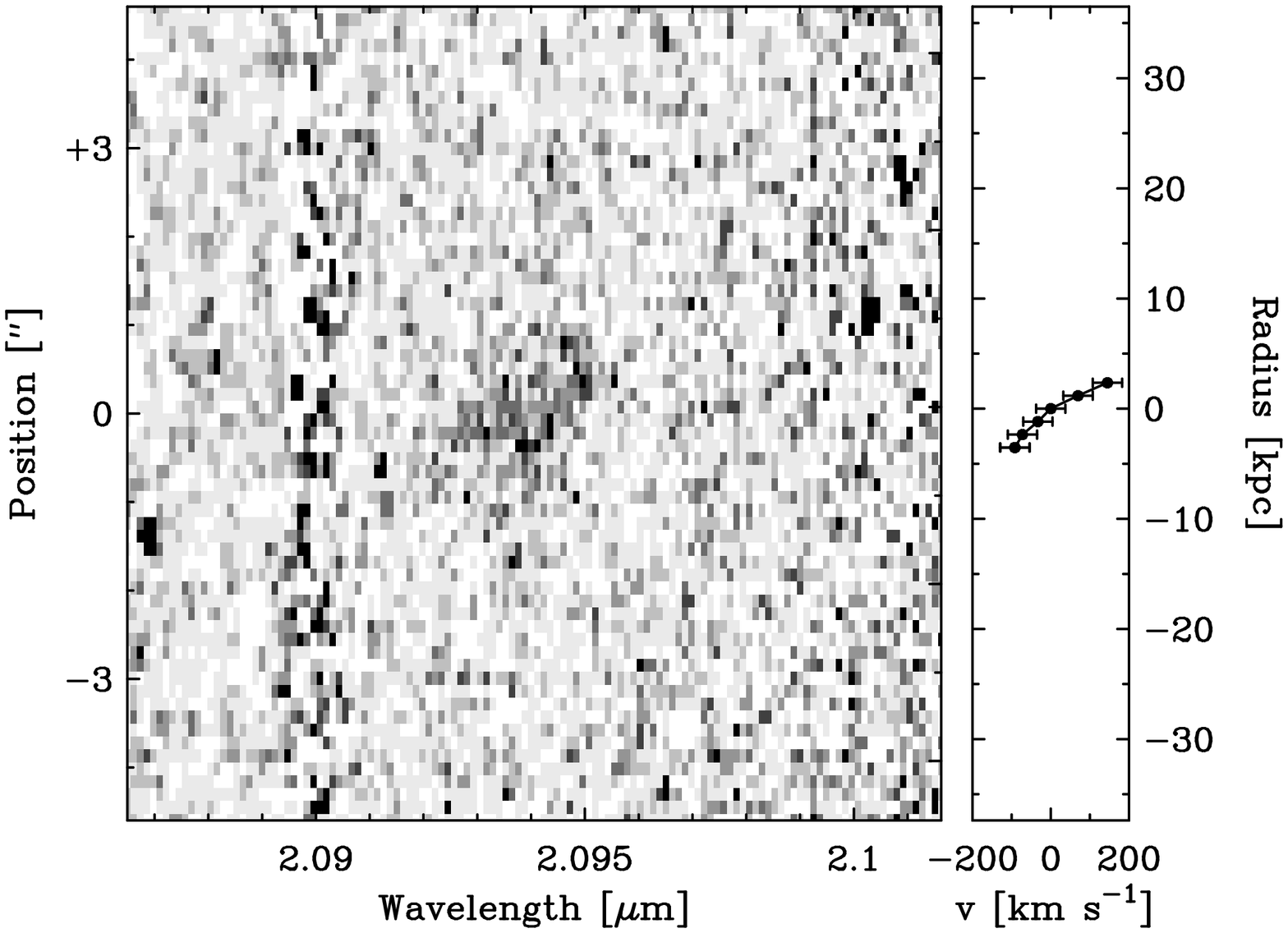}
\caption{Long-slit $\Ha$ spectrum of a galaxy at $z\sim2.2$ showing
the signature of ordered rotation (left panel); the implied rotation
curve is shown in the right-hand panel (\protect{\cite{Moorwoodetal00a}}).}
\label{fig.rcurv}
\end{figure}

Since the effects of evolution should be largest at high redshift,
$\Ha$ dynamics is particularly important for the most distant galaxies
from the surveys described in \secref{sec.surveys}. One of the
galaxies from the $z\sim2.2$ SOFI/ISAAC survey (\cite{Moorwoodetal00a})
actually shows a tilt in the 2-dimensional spectrum (taken along a
position angle within $10\deg$ of the major axis of the galaxy) 
indicating ordered
rotation (\figref{fig.rcurv}). With the short integration time of 1~hour
only than the central portion of the rotation curve is seen, so that only
a lower limit to the total line width can be determined. The resulting
peak-to-peak velocity width is $247\pm30\kms$, which increases to
$275\pm30\kms$ after applying a correction for inclination. Thus a
rotational velociy of $138\pm15\kms$ is implied at a radius of
$3\un{kpc}$. These parameters are comparable to those of local disk
galaxies which have rotation curves flattening at velocities between 100 and
$300\kms$ at radii of 1 to $5\un{kpc}$. It is therefore evident that
well-developed, relatively massive systems already occur at
$z\sim2.2$. More interestingly, with a rest-frame absolute $B$
magnitude $M_B=-22.4$ this object lies about 3 magnitudes above the
local $B$-band TF-relation. This conclusion can only be avoided if the full
rotation curves extends to over $1000\kms$ and only flattens at a
radius larger than $12\un{kpc}$ which would imply an {\it extremely\/}
massive galaxy and is therefore unlikely.

\section{Conclusions and outlook}

Surveys of $\Ha$ emission are now beginning to probe the cosmic star
formation history out to $z\sim2.2$ in a meaningful manner. However,
this field is still in a very early stage of development. A
fundamental problem is that the samples are small, leading to
uncertainty in the derived luminosity function. This is best illustrated by
the fact that even at $z\sim0$, where the available samples are
largest, significant discrepancies remain (cf., \secref{sec.z0}). 
Great progress in determining the $\Ha$
luminosity density at low redshifts should be made as a result of the
Sloan Digital Sky Survey and associated follow-up. 
At higher redshifts, the VIRMOS survey at the ESO Very Large Telescope
should provide a census out to redshifts beyond 1. Redshifts $z>2$ 
bring $\Ha$ into the $K$-band and an efficient survey would
require a cryogenic multi-object spectrograph.
At $z>2.5$ $\Ha$ shifts into the thermal infrared and
the Next Generation Space Telescope will provide the first opportunity
for deep $\Ha$ measurements at these redshifts.

As important as the composition of good samples is a good estimate of the
extinction. $\Ha$ extinctions can be determined from the Balmer
decrement, and should be feasible for most galaxies where $\Ha$ can be
detected. Furthermore, for a reliable determination of star formation
based on $\Ha$, other spectral features should also be observed 
(\cite{CharlotLonghetti01}).

First results on $\Ha$ kinematics at high redshift demonstrate the
feasibility of determining rotation curves of high-$z$ galaxies with
present-day instrumentation. Compiling a sample of high-quality $\Ha$
rotation curves at a number of redshifts out to $z\sim2.2$ will be
extremely valuable for our understanding of the disk galaxy formation
and the origin of the Tully-Fisher relation. Such a project should
ideally make use of the multiplexing capabilities of future
cryogenic multi-integral-field-unit near-infrared
spectrographs. However, a first sample of well-measured rotation
curves out to $z\sim2.2$ can already be built up with deep
near-infrared $\Ha$ spectroscopy of galaxies from the
existing samples. This is within reach with present-day near-infrared
spectrographs on large telescopes.

\begin{acknowledgments}
PvdW would like to thank the organisers for a very enjoyable meeting. It was
an honour to be able to discuss the cosmos in the same city where man first
realized that nature lends itself to analysis by the human mind. Special
thanks go to Padeli Papadopoulos and his family for excellent Greek
hospitality.
\end{acknowledgments}

{

\bibliographystyle{apalike}
\chapbblname{vdwerf}
\chapbibliography{%
strings,%
dampedLya,%
dust,%
galaxies,%
galaxyevolution,%
Hasurveys,%
HDFN,%
LBGs,%
Lyasurveys,%
PHL957,%
Q0000-2619,%
SFhistory,%
spirals,%
starbursts,%
submmsurveys%
}

\end{document}